\documentclass[12pt]{iopart}
\usepackage{iopams}
\usepackage{graphicx,color}

\newcommand{\sortsc}{\ensuremath{\tau_{\rm{o}}}}
\newcommand{\sort}{\ensuremath{T_{\rm{o}}}}

\begin{document}

\title[Number of collision and ordering time in the Jepsen gas] {Statistics
  of the total number of collisions and the ordering time in a freely
  expanding hard-point gas}

\author{Sanjib Sabhapandit$^1$, Ioana Bena$^2$ and Satya N. Majumdar$^1$}

\address{$^1$ Laboratoire de Physique Th\'eorique et Mod\`eles
Statistiques,
Universit\'e de Paris-Sud, CNRS UMR 8626, 91405 Orsay
Cedex, France \\
$^2$ Department of Theoretical Physics, University of Geneva, 
CH-1211 Geneva 4, Switzerland}

\begin{abstract}

We consider a Jepsen gas of $N$ hard-point particles undergoing free
expansion on a line, starting from random initial positions of the particles
having random initial velocities. The particles undergo binary elastic
collisions upon contact and move freely in-between collisions.  After a
certain ordering time $\sort$, the system reaches a ``fan'' state where all
the velocities are completely ordered from left to right in an increasing
fashion and there is no further collision. We compute analytically the
distributions of (i) the total number of collisions and (ii) the ordering
time $\sort$. We show that several features of these distributions are
universal.

\end{abstract}


\maketitle

\noindent\rule{\hsize}{2pt}
\tableofcontents
\noindent\rule{\hsize}{2pt}
\markboth{Number of collision and ordering time in the Jepsen gas}{}

\section{Introduction}

Exactly solvable models of interacting particle systems are important both
in equilibrium and out-of-equilibrium statistical mechanics.  Such solvable
models, apart from their pedagogical virtues, also offer important insights
and intuitions of the underlying more complex physical phenomena. Besides,
they also allow to understand better the limitations of the approximate
methods one generally uses in treating many-particles systems (e.g., the
Boltzmann equation).

One such very useful and instructive model was introduced a few decades
ago~\cite{frisch,teramoto} and is usually referred to as the {\em Jepsen
  gas}~\cite{jepsen}. It consists of identical hard-point particles that
undergo binary elastic collisions on a line.  At these instantaneous
collisions the particles exchange their velocities, while in-between
collisions they move freely.  Due to the simplicity of the dynamics, this
system admits analytical treatments for various imposed conditions, and
therefore triggered quite a lot of interest in the past.  The work of
Jepsen~\cite{jepsen} was followed by Lebowitz et
al.~\cite{lebowitz1,lebowitz2,lebowitz3}, McKean~\cite{mckean}, and
Keyes~\cite{keyes} who refined and extended the calculations of the
stochastic properties of a `test' gas particle, including its asymptotic
diffusive-like behavior and a comparison with the results of the Boltzmann
equation approach. There was a recent surge of interest in this model in
the context of the ``adiabatic piston
problem"~\cite{lieb,gruber,jarek1,jarek2,jarek3}, including the description
of the stochastic dynamics of the piston-particle~\cite{bala1} and the
analytic calculation of the energy transfer and heat flux throughout an
inhomogeneously-prepared, out-of-equilibrium configuration of the
gas~\cite{bala2}. Later on, the {\em Jarzynski theorem}~\cite{Jarzynski}
was illustrated for the case of uniform expansion or compression of the
gas~\cite{ioana}.  Jepsen gas also turned out to be useful in the study of
spin transport processes in the nonlinear
$\sigma$-model~\cite{sachdev1,sachdev2}.

In the context of biological evolution, a simplified version of Eigen's
quasispecies model~\cite{eigen}, namely, the {\em shell
  model}~\cite{krug,jain1,jain2} corresponds to the {\em free expansion} of
a Jepsen gas of $N$ particles, where initially there is one particle at
each position $-k$ for $k=1,2,\ldots,N$ with a positive velocity $U_k$
drawn independently from a position dependent probability density function
(pdf) $\phi_k(U_k)$.  A further simplification treating the velocities as
independent and identically distributed (i.i.d.)  random variables with a
common $k$ independent pdf (i.e., $\phi_k\equiv \phi\,,\forall k$) leads to
the i.i.d.  shell model, for which several asymptotic properties can be
computed exactly~\cite{satya1,ioana07}. In particular, the statistical
properties of the piston-particle (the rightmost particle or the leading
genotype in the biological language) exhibit {\em universal
  properties}. Its velocity distribution function at intermediate times
($1\ll t\ll N^\gamma$, where $\gamma$ is related to the tail of the
velocity distribution) has universal scaling behavior of only three
varieties, depending exclusively on the tail of
$\phi(U)$~\cite{ioana07}. The associated scaling functions are different
from the usual extreme-value distribution of uncorrelated random variables,
and this difference is due to the dynamically-built, collision-generated
correlations between the particles at finite times. These correlations are
also responsible for the fact that the statistics of the piston collisions
is not Poissonian. Indeed, both the mean and the variance of the number of
collisions the piston-particle undergoes increase logarithmically with
time, but with different prefactors (that are also universal). \\

\begin{figure}
\centering \includegraphics[width=.95\textwidth]{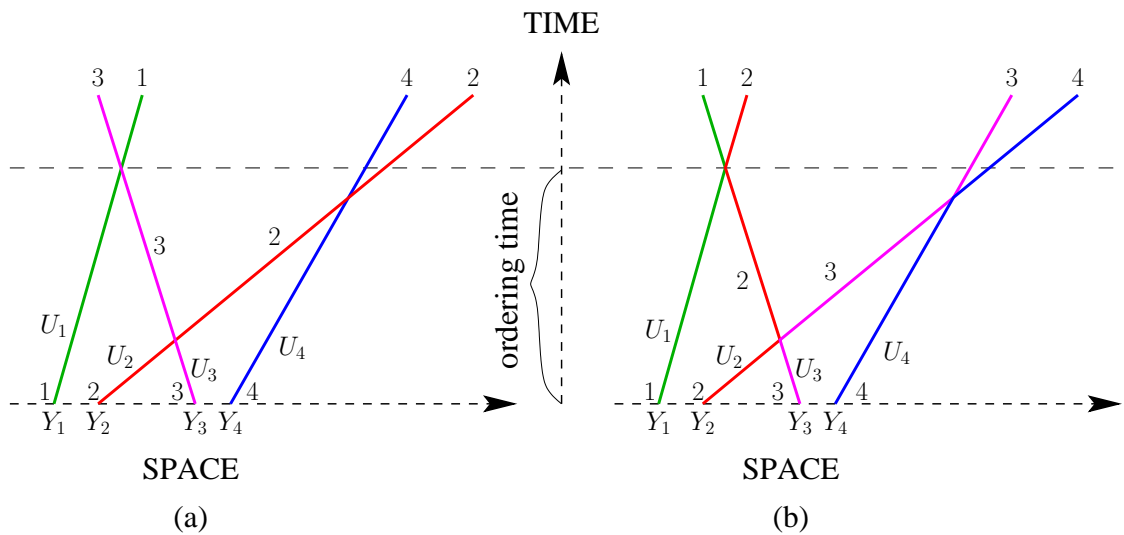}
\caption{(Color online).  (a) A realization of free trajectories for $N=4$
  particles.  The trajectories are labeled according to the order of their
  starting positions from left to right, $Y_1<Y_2<Y_3<Y_4$.  The slopes
  $\{U_1, U_2, U_3,U_4\}$ of the trajectories with respect to the
  ``TIME''-axis represent velocities associated with them.  (b) Actual
  particle-trajectories.  After a ordering time the system reaches a
  ``fan'' state where the velocities are completely ordered from left to
  right in an increasing fashion and there is no further collision.  The
  total number of binary collisions $N_c=3$}
\label{traj}
\end{figure}

Despite this rather rich history of the Jepsen gas, there are other
interesting natural questions to which detailed answers are missing.
Consider, for example, the fact that the dynamics of the freely expanding
Jepsen gas provides a ``natural" sorting algorithm for the velocities of
the particles. It leads to an asymptotic ``fan'' state (see \fref{traj}) in
which these velocities are completely ordered from left to right in an
increasing fashion.  Two questions arise naturally from this ordering of
the Jepsen gas are
\begin{enumerate}
\item What is the total number of collisions the particles undergo?
\item How long does it take for the gas to reach the ``fan'' state?
\end{enumerate}
In this paper we provide analytical answers to these two questions.

More precisely, we consider the free expansion of a Jepsen gas of $N$
hard-point particles of equal mass on the infinite real line
$(-\infty,\infty)$.  The initial positions of the particles are drawn
independently from a common pdf $\psi(X)$ and a random velocity drawn
independently from a common pdf $\phi(U)$ is assigned to each particle.  At
subsequent times ($t>0$) each particle moves ballistically according to its
assigned velocity, and upon contact between two particles they undergo
elastic collision which merely interchanges their respective velocities.
We assume both the set of positions and the set of velocities are
continuous variables (or at least one set), so that the collisions are
always binary, and there can be at most one binary collision at one instant
of time. Clearly, in each collision the velocities of the colliding
particles get ordered such that after the collision the particle on the
right acquires the larger of the two velocities and the one on the left
gets the smaller of the two velocities.  Therefore, after a certain
ordering time the system reaches a ``fan'' state where the velocities of
the particles are increasingly ordered from left to right. Once this
``fan'' state is reached, evidently, there cannot be any further collision.
For any given initial realization of positions and velocities, the dynamics
of future evolution of the gas is completely deterministic. Therefore, both
the total number of collisions (denoted by $N_c$) and the ordering time
(denoted by $\sort$) are solely determined by the initial condition. Hence,
$N_c$ and $\sort$ are random variables in the sense that they differ from
one realization of initial condition to another.  In this paper we
analytically compute their distributions.
Our main results are:
\begin{enumerate}

\item The probability $P_N(N_c)$ of having $N_c$ collision, is completely
  independent of $\psi(X)$ and $\phi(U)$ for all $N\geqslant 2$, and for large
  $N$ it approaches a Gaussian form around its mean $\langle N_c\rangle = N
  (N-1)/4$ with a variance $\langle N_c^2\rangle-\langle N_c\rangle^2 =
  N(N-1)(2N+5)/72$.

\item When the ordering time $\sort$ is suitably scaled with $N$ as
  $\sortsc=\sort/[bN^2]$, ---where $b$ is a {\em nonuniversal} scale factor
  which depends explicitly on $\psi(Y)$ and $\phi(U)$ as given by
  \eref{expression of b},--- the limiting pdf of $\sortsc$ in the scaling
  limit $N\rightarrow\infty$, $\sort\rightarrow\infty$ while keeping
  $\sortsc$ fixed, becomes {\em universal}, i.e., completely independent of
  $\psi(X)$ and $\phi(U)$, and is given by the well known Fr\'echet form
  $f(\sortsc)={\sortsc}^{-2}\,\exp(-1/\sortsc)$ that arises in the extreme
  value statistics.

\end{enumerate}

The paper is organized as follows.  In~\sref{collision}, we compute the
statistics of the total number of collisions, and
in~\sref{section_sorting_time}, we compute the limiting distribution of the
ordering time.  \Sref{remarks} contains some concluding remarks. Most of
the calculational details are relegated to the Appendices A--C.

\section{Total number of collisions}
\label{collision}

In order to express the total number of collisions $N_c$ in terms of the
initial condition, we label the particles as $i=1,2,\ldots,N$ from left to
right, i.e., $i=1$ is the leftmost particle and $i=N$ is the rightmost one
(see \fref{traj}). Let $Y_i$ for each $i=1,2,\ldots,N$ denote the initial
position of the $i$-th particle such that $Y_1<Y_2<\cdots<Y_N$.  Note that
the labeled, ordered coordinates $\{Y_i\}$'s are no longer distributed
independently according to $\psi(Y_i)$, rather their joint pdf is given by
\begin{equation}
\psi_{\rm{joint}}(Y_1,Y_2,\ldots,Y_N) =N!\prod_{i=1}^N \psi(Y_i)
\prod_{j=1}^{N-1}\theta(Y_j-Y_{j+1}),
\end{equation}
where $\theta(x)$ is the Heaviside step function. On the other hand, the
initial velocities associated with the labeled particles, which we denote
by $U_i$ for $i=1,2,\ldots,N$ are i.i.d. random variables drawn from the
common pdf $\phi(U)$.

Now, for a given initial configuration, the system is fully and uniquely
described at all subsequent times by the set of the free trajectories
$\{Y_{k}+U_{k} t\,|\;k=1,2,\ldots,N\;\mbox{and}\; t\geqslant 0\}$ (see
\fref{traj}(a)).  {\em Note that the $k$-th free trajectory should not be
  confused with the actual trajectory of the $k$-th particle}.  Indeed,
each of the particles travels along such a free trajectory until it
collides with another particle; in such a binary collision the particles
interchange their trajectories (see \fref{traj}(b)).  In terms of the free
trajectories, the total number of collisions $N_c$ is just the total number
of intersections among the $N$ free trajectories.  Two free trajectories
can, of course, intersect at most once.

Consider first two extreme situations.  If in the initial configuration the
velocities are already in the increasing order, i.e., $U_1<U_2< \cdots
<U_N$ (which happens with probability $1/N!$), then there cannot be any
collision at subsequent times as the gas evolves. Thus, in this case, the
system is always in the ``fan'' state and $N_c=0$. On the other hand, if
the initial velocities are in the decreasing order, i.e., $U_1>U_2> \cdots
>U_N$ (which also happens with probability $1/N!$), then each pair of free
trajectories intersects once and there are ${N\choose 2}$ of them.
Therefore, this configuration yields the maximum number of binary
collisions, which is $N_c=N(N-1)/2$.  For any other realizations of the
velocities, $N_c$ lies between $0$ and $N(N-1)/2$.  The total number of
collisions $N_c$ is a random variable, $N_c\in[0,N(N-1)/2]$, which differs
from one realization of the initial configuration of the particle
velocities to another, and whose statistical properties we address below.

Since, $Y_i<Y_j$ for $i<j$, two free trajectories $(Y_i+U_i t)$ and
$(Y_j+U_j t)$ intersect if and only if $U_i>U_j$ for $i<j$.  Therefore, for
a given realization of the initial velocities $\{U_i\}$, the total number
of collisions (total number of intersections among $N$ free trajectories)
can be expressed as
\begin{equation}
N_c=\sum_{i=1}^{N-1}\sum_{j=i+1}^{N} \theta(U_i-U_j).
\label{N_c}
\end{equation}
Obviously, $N_c$ is independent of the set of initial positions $\{Y_i\}$
of the particles and their distribution. Moreover, we will show below that
$P_N(N_c)$ is also completely {\em independent of the velocity
  distribution} $\phi(U)$ and is solely determined by $N$.

The probability $P_N(N_c)$ of having $N_c$ total number of collisions can
be formally expressed as
\begin{eqnarray}
P_N(N_c)=\displaystyle\int \cdots \displaystyle\int 
\delta\left[ N_c-\sum_{i=1}^{N-1}\sum_{j=i+1}^{N}
\theta(U_i-U_j) \right]\,
\prod_{i=1}^N \phi(U_i) \,\rmd U_i
\label{prob1}
\end{eqnarray}
where $\delta[n]$ with integer $n$, is discrete delta function:
$\delta[0]=1$ and $\delta[n]=0$ for any $n \not=0$.  Let us consider the
change of variables~\cite{satya06}
\begin{equation}
u_i=\int_{-\infty}^{U_i} \phi(U)\,\rmd U\,, \quad \mbox{i.e.,} \quad \rmd
u_i= \phi(U_i) \,\rmd U_i\quad \mbox{for}~~i=1,...,N.
\label{change-variables}
\end{equation}
Obviously $u_i$ is a monotonically increasing function of $U_i$, and therefore
$\theta(U_i-U_j)=\theta(u_i-u_j)$. Moreover, since $\phi(U)$ is normalized
to unity, the variables $u_i$ vary from 0 to 1, and so \eref{prob1} becomes
\begin{equation}
P_N(N_c)=\int_0^1\rmd y_1\dots \int_0^1\rmd y_N\; \delta\left[
  N_c-\sum_{i=1}^{N-1}\sum_{j=i+1}^{N} \theta(u_i-u_j) \right],
\label{prob2}
\end{equation}
where the velocity distribution $\phi(U)$ simply drops out, meaning that
$P_N(N_c)$ is {\em universal}, i.e., is the same as if the ``new"
velocities $\{u_i\}$ are drawn independently from an uniform distribution
over $[0,1]$.

The mean number of total collisions $\langle N_c\rangle$ is straightforward
to compute.  Using the change of variables in \eref{change-variables}, it
is trivially checked that $\langle \theta(U_i-U_j)\rangle=1/2$. Therefore,
taking the average in \eref{N_c} yields
\begin{equation}
\langle N_c\rangle = \sum_{i=1}^{N-1}\sum_{j=i+1}^{N} \langle
\theta(U_i-U_j)\rangle=\frac{N (N-1)}{4}\sim \frac{N^2}{4} ~~\mbox{as}~
N\rightarrow\infty.
\label{mean}
\end{equation}

The calculation of the variance implies several steps that are indicated
in~\ref{varianceap}. One obtains finally:
\begin{equation}
\sigma^2= \langle N_c^2\rangle-\langle N_c\rangle^2 =
\frac{N(N-1)(2N+5)}{72}\sim\frac{N^3}{36} ~~\mbox{as}~ N\rightarrow\infty.
\label{variance}
\end{equation}

\begin{figure}
\centering \includegraphics[width=0.5\columnwidth]{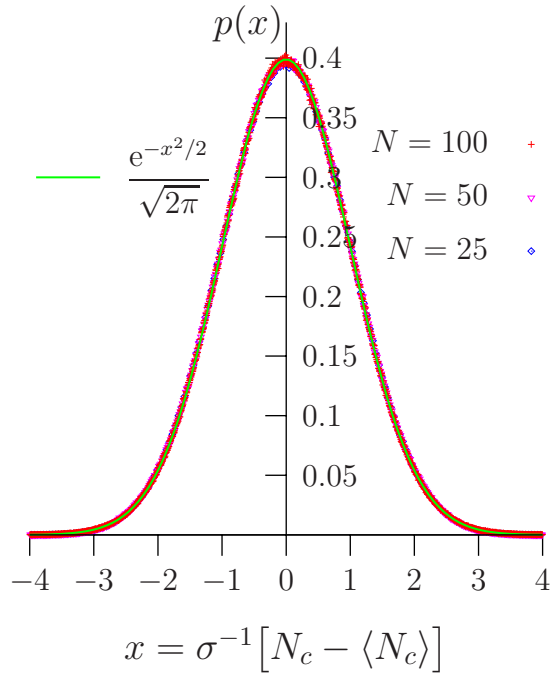}
\caption{(Color online). Probability density of the total number of
  collisions (scaled by the mean and the variance) for $N=25,\, 50$, and
  $100$ ({\color{blue}\opendiamond}, {\color{magenta}\opentriangledown},
  and {\color{red}+} respectively as symbols), as compared to the Gaussian
  $p(x)=\exp(-x^2/2)/\sqrt{2\pi}$ (drawn with {\color{green}\full}).  We
  scaled the numerically-obtained data by the mean $\langle N_c\rangle$ and
  the variance $\sigma^2$ that are given as functions of $N$ by \eref{mean}
  and \eref{variance} respectively. It is difficult to notice any
  difference between the three data sets and the Gaussian.}
\label{distribution}
\end{figure}

As given by \eref{N_c}, $N_c$ is a sum of $N(N-1)/2$ random variables that
are {\em correlated}. However, each term in \eref{N_c} is correlated with
only $2(N-2)$ other terms, since two different terms are correlated only
when they have one velocity in common.  Therefore, near the mean $\langle
N_c\rangle$ and within a region $\Or(\sigma)$, the probability distribution
$P_N(N_c)$ for large $N$ has a Gaussian form (see \ref{PDF N_c}),
\begin{equation}
P_N(N_c)\approx \frac{1}{{\sqrt{2\pi}\sigma}}\; \exp\left(-\frac{[N_c-\langle
  N_c\rangle]^2}{2\sigma^2}\right).
\end{equation}
One can check numerically that this is actually already an extremely good
approximation for $N$ as small as $25$ and for practically all the values
of $N_c$, as illustrated by the \fref{distribution}.

\section{Ordering  time}
\label{section_sorting_time}

\begin{figure}
\centering \includegraphics[width=.7\textwidth]{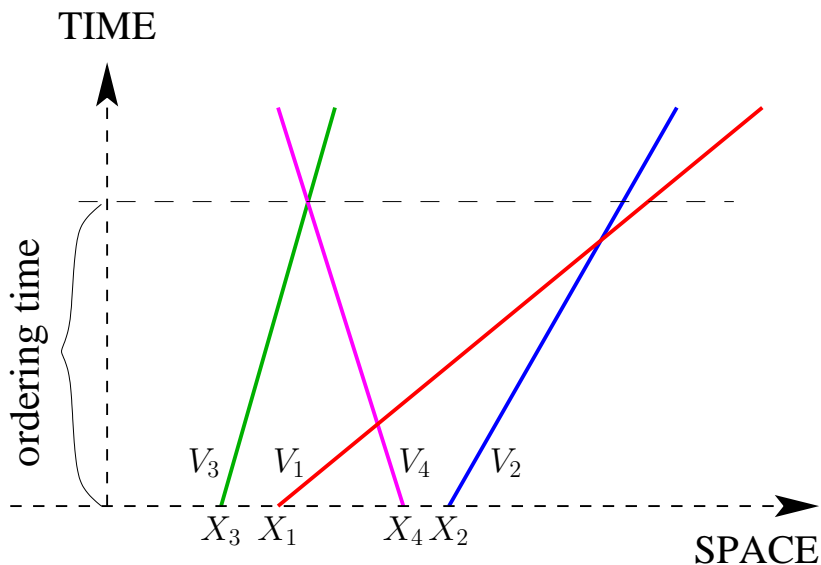}
\caption{(Color online).  The same realization of the free trajectories
  shown in \fref{traj}. However, now the trajectories are labeled according
  to the decreasing order of the velocities $V_1>V_2>V_3>V_4$ and the
  initial positions $\{X_1,X_2,X_3,X_4\}$ are indexed according to the
  order of the velocities.  Note $V_1=U_2, V_2=U_4, V_3=U_1, V_4=U_3$ and
  $X_1=Y_2, X_2=Y_4, X_3=Y_1, X_4=Y_3$, while comparing with \fref{traj}.}
\label{traj2}
\end{figure}

In order to compute the statistics of the ordering time $\sort$, it is
convenient to relabel the particles according to the decreasing order of
the velocities (see \fref{traj2}). Let $\{V_1,V_2,\ldots,V_N\}$ denote the
decreasingly ordered set of the initial velocities, i.e.,
\begin{eqnarray*}
V_1&=\max(U_1,U_2,\ldots,U_N),\\
V_2&=\max(\{U_1,U_2,\ldots,U_N\}\setminus\{V_1\}),\\
V_3&=\max(\{U_1,U_2,\ldots,U_N\}\setminus\{V_1,V_2\}),\\
&\vdots\\
V_N&=\min(U_1,U_2,\ldots,U_N),
\end{eqnarray*}
so that $V_1>V_2>\cdots>V_N$. The joint pdf of these ordered velocities is
given by
\begin{equation}
\phi_{\rm{joint}}(V_1,V_2,\dots,V_N)=\left[N!\prod_{i=1}^N
  \phi(V_i)\right]\times \left[\prod_{i=1}^{N-1}\theta(V_i-V_{i+1})\right].
\label{joint velocity distribution}
\end{equation}
Let $X_i$ for each $i=1,2,\ldots,N$ denote the initial position of the
particle having the initial velocity $V_i$.  Note that when we label the
particles according to the order of their velocities, their initial
positions $\{X_i\}$'s are no longer ordered on the line, but are
i.i.d. random variables drawn from the pdf $\psi(X)$.

As before, for a given initial condition, the system is fully and uniquely
described at all subsequent times by the set of free trajectories
$\{X_i+V_i t\,|\; i=1,2,\ldots,N\;\mbox{and}\; t\geqslant 0\}$, where
$V_1>V_2>\dots>V_N$.  The ``fan'' state is reached when all the free
trajectories become completely ordered according to the velocities, i.e.,
$X_1+V_1t>X_2+V_2t>\cdots>X_N+V_Nt$ at all $t>\sort$ (see \fref{traj2}).

Let $F_N (T)$ be the cumulative probability distribution of the ordering
time, i.e., $F_N(T)=\mathrm{Prob}[\sort<T]$.  It immediately follows that
$F_N(T)= \mathrm{Prob}[X_1+V_1T>X_2+V_2T>\cdots>X_N+V_NT]$, given that
$V_1>V_2>\dots>V_N$. This probability can be formally expressed as
\begin{equation}
F_N(T)=\left\langle \prod_{i=1}^{N-1}
\theta\Bigl([X_i-X_{i+1}]+T[V_i-V_{i+1}]\Bigr)\right\rangle,
\label{sorting}
\end{equation}
where $\langle\dots\rangle$ denotes the averaging over both the
i.i.d. initial positions that are drawn from the common pdf $\psi(X)$, and
the initial ordered velocities, whose joint pdf is given by \eref{joint
  velocity distribution}. Note that, $F_N(0)$ for any finite $N$ is
non-zero,
\begin{equation}
F_N(0)=\left\langle \prod_{i=1}^{N-1}
\theta\Bigl(X_i-X_{i+1}\Bigr)\right\rangle =\frac{1}{N!}
\label{sorting 0}
\end{equation}
and is universal.  This represents the probability that the initial
positions of the particles are also ordered according to their velocities,
i.e., $X_1>X_2>\cdots>X_N$, so that there cannot be any collision. The pdf
of the ordering time is simply
\begin{equation}
f_N(\sort)=\frac{\delta(\sort)}{N!} + \frac{\rmd F_N(\sort)}{\rmd \sort},
\end{equation}
where the extra weight at $\sort=0$ vanishes in the limit
$N\rightarrow\infty$.

Let us introduce the variables $x_i=X_{i+1}-X_{i}$ and $v_i=V_i-V_{i+1}$
for brevity. In terms these variables we rewrite the
expression~\eref{sorting} as
\begin{equation}
F_N(T)=\left\langle \prod_{i=1}^{N-1}
\Biggl[1-\theta(x_i)\, 
\theta\Bigl(\frac{x_i}{T}-v_i\Bigr)
\Biggr]
\right\rangle,
\label{sorting-2}
\end{equation}
where $v_i>0$ for $i=1,2,\ldots,N-1$.

\begin{figure}
\centering \includegraphics[width=.8\textwidth]{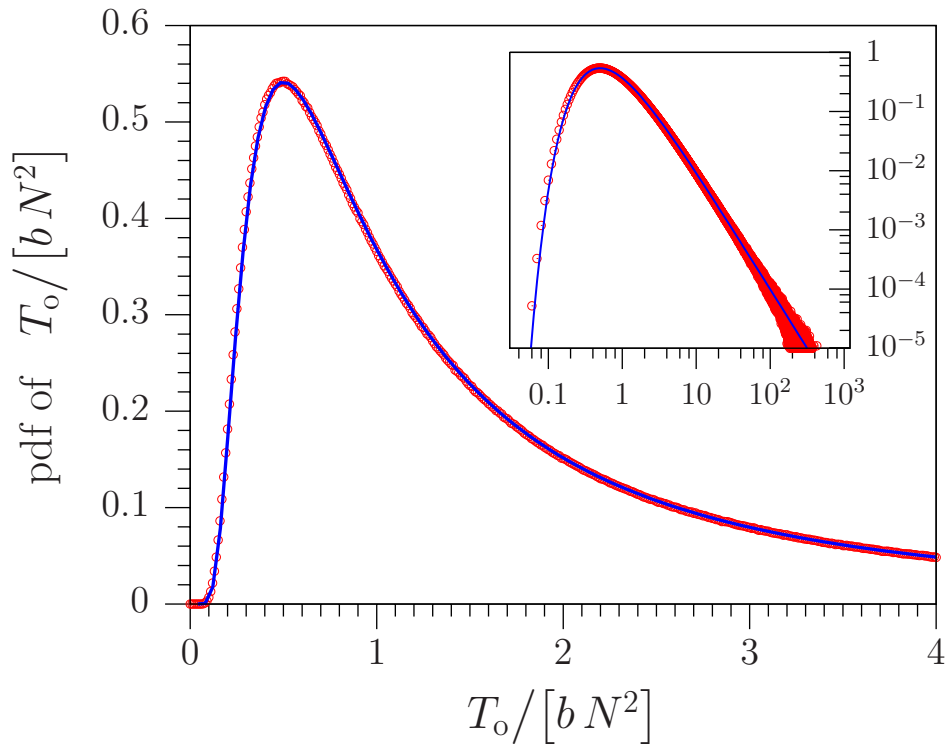}
\caption{(Color online).  The pdf of the ordering time $\sort$, plotted
  using the scaled variable $\sortsc=\sort/[b\, N^2]$, where $b$ is given
  by \eref{expression of b}. The points ({\color{red} $\odot$}) were
  obtained by numerical simulation with $N=1000$, averaging over $10^8$
  realizations of initial configurations where particles were distributed
  uniformly and independently in the interval $[0,1]$, and the velocities
  were drawn independently from a Gaussian pdf
  $\phi(V)=\exp(-V^2/2)/\sqrt{2\pi}$. \Eref{expression of b} gives
  $b=1/\bigl[12\,\sqrt{\pi}\bigr]$.  The solid line
  ({\color{blue}\rule[.5ex]{7mm}{1pt}}) represents the Fr\'echet pdf
  $f(\sortsc)={\sortsc}^{-2}\,\exp(-1/\sortsc)$.  The inset displays the
  same pdf-s on a logarithmic scale.}
\label{frechet}
\end{figure}

Since $\{X_i\}$'s are drawn independently from $\psi(X)$, two different
variables $x_i$ and $x_j$ with $i\not=j$ are correlated iff
$|i-j|=1$. Therefore, for large $N$, treating $\{x_i\}$'s as i.i.d. random
variables, with a common pdf $g(x)=\displaystyle\int_{-\infty}^{\infty}
\psi(X)\, \psi(X+x) \,\rmd X$, is a rather good approximation, which
becomes exact in the limit $N\rightarrow\infty$.  On the other hand, in the
variables $v_{i}=V_i-V_{i+1}$ ($i=1,2,\ldots,N-1$), the ordered velocities
$\{V_i\}$ are correlated, as can be seen from their joint pdf given by
\eref{joint velocity distribution}. Therefore, the random variables
$t_i=\theta(x_i) [x_i/v_i]$ with $i=1,2,\ldots,N-1$, are not independent
either. It turns out, however, that the limiting distribution of the
ordering time $\sort=\max(t_1,t_2,\ldots, t_{N-1})$, when suitably scaled
with $N$, has the well known Fr\'echet form (see \fref{frechet}) that
arises in the extreme value statistics of i.i.d. random variables drawn
from a common parent distribution with a power-law tail.  To show this, we
formally expand the product in \eref{sorting-2} as a sum of terms
\begin{eqnarray}\fl\qquad
F_N(T)=1 &-
\sum_{i=1}^{N-1}\left\langle \theta(x_i)\, 
\theta\Bigl(\frac{x_i}{T}-v_i\Bigr) \right\rangle\nonumber\\
&+\sum_{i=1}^{N-2}\sum_{j=i+1}^{N-1}\left\langle \theta(x_i)\theta(x_j)\,  
\theta\Bigl(\frac{x_i}{T}-v_i\Bigr) 
\theta\Bigl(\frac{x_j}{T}-v_j\Bigr) \right\rangle 
\;-\cdots\nonumber\\
&+(-1)^n\sum_{i_1=1}^{N-n}\;
\sum_{i_2=i_1+1}^{N-n+1} 
\cdots\sum_{i_n=i_{n-1}+1}^{N-1}
\left\langle
\prod_{\nu=1}^n\Biggl[ 
\theta(x_{i_\nu})\, 
\theta\Bigl(\frac{x_{i_\nu}}{T}-v_{i_\nu}\Bigr)\Biggr]
 \right\rangle \nonumber\\
&+\cdots
\label{expansion of F_N}
\end{eqnarray}

We show in~\ref{terms of F_N} that $T$ scales as $N^2$ when both $T$ and
$N$ are large.  Moreover, as detailed in ~\ref{terms of F_N}, considering
first $n\ll N$, and then taking the scaling limit $T\rightarrow\infty$ and
$N\rightarrow\infty$, but keeping $T/N^2$ fixed, the $n$-th term (counting
from the ``zero''-th term, which equals unity) in \eref{expansion of F_N}
becomes
\begin{equation}\fl\qquad\quad
\lim_{N\rightarrow\infty}
\sum_{i_1=1}^{N-n}\;
\sum_{i_2=i_1+1}^{N-n+1} 
\cdots\sum_{i_n=i_{n-1}+1}^{N-1}
\left\langle
\prod_{\nu=1}^n\Biggl[ 
\theta(x_{i_\nu})\, 
\theta\Bigl(\frac{x_{i_\nu}}{N^2\tau'}-v_{i_\nu}\Bigr)\Biggr]
 \right\rangle = \frac{1}{n!} \left(\frac{b}{\tau'}\right)^n,
\end{equation}
where 
\begin{equation}
b=\left[\int_0^\infty x\, g(x)\,\rmd x \right] \times
\left[\int_{-\infty}^\infty \phi^2(V)\,\rmd V \right]\,,
\label{expression of b}
\end{equation} 
with, recall,
\begin{equation}
g(x)=\int_{-\infty}^{\infty} \psi(X)\, \psi(X+x) \,\rmd X.
\label{expressiong}
\end{equation}
Thus, \eref{expansion of F_N} yields
\begin{equation}
\lim_{N\rightarrow\infty} F_N(b\,N^2\,\tau)=\sum_{n=0}^\infty
\frac{(-1)^n}{n!} \left(\frac{1}{\tau}\right)^n
=\exp\left(-\frac{1}{\tau}\right).
\label{frechet 2}
\end{equation}
The limiting pdf of the scaled ordering time $\sortsc=\sort/[b\, N^2]$,
---where the scale factor $b$ is {\em nonuniversal} and is given explicitly
by \eref{expression of b},--- is therefore given by the {\em universal}
function $f(\sortsc)={\sortsc}^{-2}\,\exp(-1/\sortsc)$.  \Fref{frechet}
compares the results of the numerical simulation of the gas with this
limiting pdf.

The Fr\'echet cumulative distribution \eref{frechet 2} usually emerges
as the cumulative distribution of the maximum of a set of $N^2/2$
i.i.d. random variables $\{\tau_i\}$ each drawn from the common pdf having a
power-law tail $p(\tau)\sim 2 b/\tau^2$. In the context of Jepsen gas, let us
consider two free trajectories $(X_i+V_i t)$ and $(X_j+V_j t)$ chosen at
random. The ordering time $\tau_{ij}$ between these two free trajectories
has the pdf 
\begin{equation}
p(\tau_{ij})=
\left\langle\delta\left(\tau_{ij}-\theta\left(\frac{X_j-X_i}{V_i-V_j}\right)
\left[\frac{X_j-X_i}{V_i-V_j}\right]
\right) 
\right\rangle
\sim\frac{2b}{\tau_{ij}^2}
~~\mbox{for large}~\tau_{ij},
\end{equation}
where $b$ is given by \eref{expression of b}. The ordering time $\sort$
of the full system is clearly the maximum of the ordering times
$\{\tau_{ij}\}$ between each of the $N(N-1)/2$ ($\approx N^2/2$ for large
$N$) pairs of free trajectories. Therefore, it turns out simply that in the
$N\rightarrow\infty$, the ordering times between different pairs of free
trajectories become uncorrelated, and can be treated as i.i.d. random
variables drawn from a common pdf having power-law tail $p(\tau)\sim 2
b/\tau^2$.

\section{Concluding remarks}
\label{remarks}

In this paper we have computed the statistics of the total number of
collisions of a freely expanding gas of $N$ hard-point particles, and found
that the mean is $\Or(N^2)$. On the other hand, the variance is $\Or(N^3)$,
due to the correlations between different particle collisions, without
which the variance would also have been $\Or(N^2)$. However, despite these
correlations, the probability distribution of the total number of
collisions near the mean is very well approximated by a Gaussian form.

In the context of biological evolution of quasispecies, the {\em evolution
  time} $T_{\rm e}$ ---which is defined as the time at which the rightmost
particle undergoes the last collision, i.e., the free trajectory with the
largest slope (with respect to the time-axis) becomes the rightmost
trajectory--- was studied recently~\cite{krug, krug1}.  It was estimated
that its pdf has a power-law tail $p(T_{\rm e})\sim T_{\rm e}^{-2}$.  The
{\em ordering time} studied here represents obviously the upper bound to
this evolution time, $T_{\rm e} \leqslant \sort$. We have found that the
limiting distribution of the ordering time, when suitably scaled with $N$,
becomes universal and is given by the Fr\'echet form. This Fr\'echet form
usually appears as the limiting distribution of maximum of a set of
i.i.d. random variables drawn from a common distribution with a power-law
tail. Our result here provides another mechanism of generating Fr\'echet
form as a limiting distribution.

Finally, while we have studied the statistics of the total number of
collisions, it would be interesting to study the number of collisions of
the particles $N_c(t)$ up to time $t$. Our results correspond to the limit
$t\rightarrow\infty$. In this context we point out that recently the
collision statistics of a tagged particle in a $d$-dimensional hard-sphere
gas at equilibrium was investigated using Boltzmann equation
approach~\cite{visco} and was found to be non-Poissonian.  It would be
interesting to verify this conclusion for a tagged particle in the 1-$d$
Jepsen gas for which an exact result may be possible to obtain.

\addcontentsline{toc}{section}{\protect\numberline{}Acknowledgments}

\ack SS and SNM acknowledge the support of the Indo-French Centre for the
Promotion of Advanced Research (IFCPAR/CEFIPRA) under Project 3404-2.

\makeatletter%
\def\@sect#1#2#3#4#5#6[#7]#8{\ifnum #2>\c@secnumdepth
  \let\@svsec\@empty\else
  \refstepcounter{#1}\edef\@svsec{\csname the#1\endcsname. }\fi
  \@tempskipa #5\relax
  \ifdim \@tempskipa>\z@
  \begingroup #6\relax
  \noindent{\hskip #3\relax\@svsec}{\interlinepenalty \@M #8\par}%
  \endgroup
  \csname #1mark\endcsname{#7}\addcontentsline
          {toc}{#1}{\ifnum #2>\c@secnumdepth \else
            \protect\numberline{}{\csname the#1\endcsname}\fi
            . #7}\else
          \def\@svsechd{#6\hskip #3\relax  
            \@svsec #8\csname #1mark\endcsname
                    {#7}\addcontentsline
                    {toc}{#1}{\ifnum #2>\c@secnumdepth \else
                      \protect\numberline{}{\csname the#1\endcsname}\fi
                      . #7}}\fi
          \@xsect{#5}}
\makeatother%

\appendix

\section{Variance of the total number of collisions}
\label{varianceap}

Subtracting the mean given by \eref{mean} from \eref{N_c}, then taking the
square and the average gives
\begin{equation}
\sigma^2=
\left\langle 
\Bigl[N_c-\langle N_c\rangle\Bigr]^2\right\rangle 
= \frac{1}{4}\sum_{i=1}^{N-1}\sum_{j=i+1}^{N}
\sum_{k=1}^{N-1}\sum_{l=k+1}^{N} \langle s_{ij}s_{kl}\rangle,
\label{N_c^2}
\end{equation}
where 
\begin{equation}
s_{ij}=2\,\theta(U_i-U_j) -1
=\pm 1.
\label{S_ij}
\end{equation} 
Note that, 
$\langle s_{ij} \rangle =0$, $\langle s_{ij}^2 \rangle =1$ and
$\langle s_{ij}s_{kl}\rangle= 4
\langle\theta(U_i-U_j)\,\theta(U_k-U_l)\rangle -1$.

\begin{figure}[ht!]
\centering \includegraphics[width=8cm]{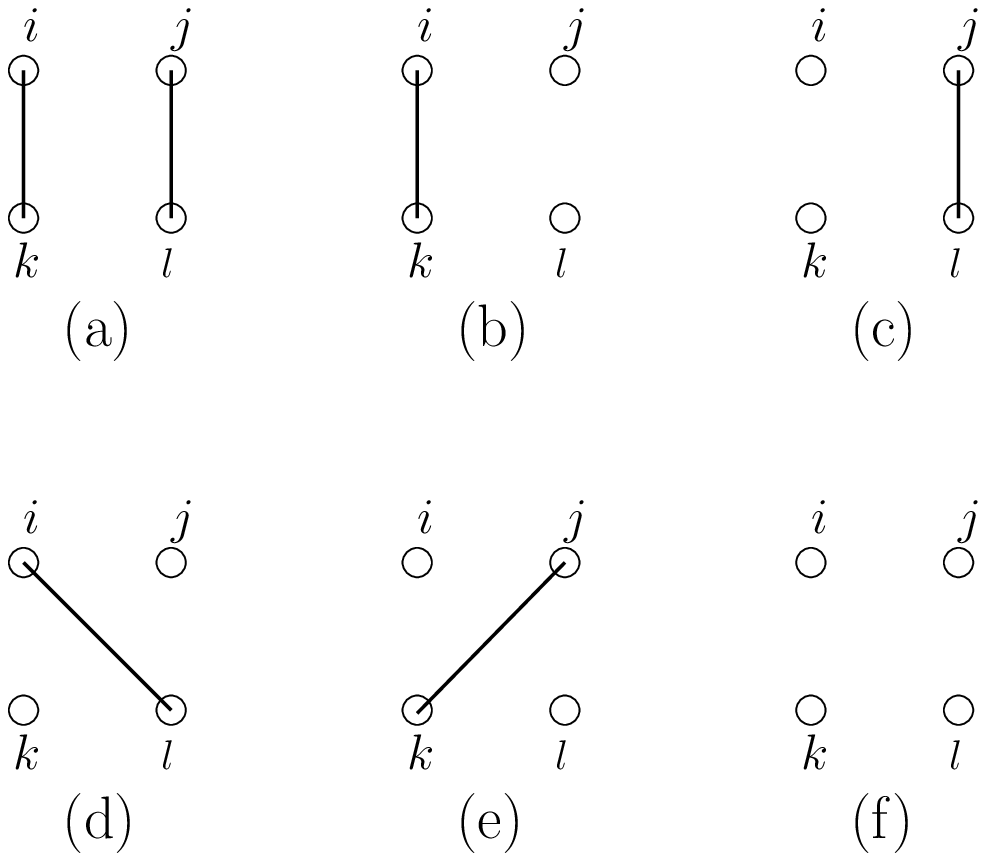}
\caption{Diagrammatic representations of the different types of terms
  appearing in \eref{N_c^2} for the computation of $\langle
  N_c^2\rangle$. In all the diagrams $j>i$ and $l>k$. The solid lines
  connect two indices having equal value. One has: (a) $i=k$ and $j=l$;
  (b) $i=k$ and $j\ne l$; (c) $i\ne k$ and $j=l$; (d) $i=l$; (e) $j=k$; (f)
  $i \ne k$ and $j\ne l$.}
\label{diagrams}
\end{figure}

The total number of terms in the summation in \eref{N_c^2} is
$[N(N-1)/2]^2$. These can be grouped according to the correlation functions
$\langle s_{ij}s_{kl}\rangle$'s that are of similar kind. This can be
conveniently represented by diagrams as shown in Fig.~\ref{diagrams}.  Then
using the change of variables~\eref{change-variables}, it is easy to
compute the correlations $\langle \theta(U_i-U_j)\theta(U_k-U_l)\rangle$
and hence $\langle s_{ij}s_{kl}\rangle$, corresponding to each of the
diagrams (a)--(f) in \fref{diagrams}.

\begin{description}

\item[Diagram (a):] $\langle s_{ij}s_{kl}\rangle \equiv \langle
  s_{ij}^2\rangle=1$.  

The number of such terms in \eref{N_c^2} is equal to number of ways of
choosing two indices out of $N$, which is
$${N\choose 2}=\frac{N(N-1)}{2}\,. $$

\item[Diagram (b):]  $\langle s_{ij} s_{kl}\rangle\equiv\langle
  s_{ij}s_{il}\rangle=1/3$.  

The number of such terms in \eref{N_c^2} is equal to number of ways of
choosing three indices $\{i,j,l\}$ out of $N$ with allowing permutation
between two of them $\{j,l\}$. This is given by
$$2\times {N \choose 3}=\frac{N(N-1)(N-2)}{3}\,.$$

\item[Diagram (c):] Same as for diagram (b).

\item[Diagram (d):] $\langle s_{ij} s_{kl}\rangle =\langle
  s_{ij}s_{ki}\rangle =-1/3$.  

The number of such terms in \eref{N_c^2} is just the number of ways of
choosing three indices $\{k,i,l\}$ out of $N$, which is
$${N \choose 3}=\frac{N(N-1)(N-2)}{6}\,.$$

\item[Diagram (e):] Same as for diagram (d).

\item[Diagram (f):] $\langle s_{ij}s_{kl}\rangle = \langle
  s_{ij}\rangle\langle s_{kl}\rangle = 0$. These terms do not
  contribute to the sum in~\eref{N_c^2}.
\end{description}

Therefore, finally \eref{N_c^2} yields
\begin{equation}
\sigma^2= \frac{1}{4}\left[1\cdot{N\choose 2} 
+ 2\cdot\frac{1}{3}\cdot{N\choose 3}\right] =
\frac{N(N-1)(2N+5)}{72}\,.
\end{equation}

\section{Probability distribution of the total number of collisions}
\label{PDF N_c}

Let us consider the deviation of the total number of collisions given by
\eref{N_c} from it mean given by \eref{mean}:
\begin{equation}
M=N_c-\langle N_c\rangle = \frac{1}{2}
\sum_{i=1}^{N-1}\sum_{j=i+1}^{N} \mathrm{sgn}(U_i-U_j)
=\sum_{i=1}^{N-1}\sum_{j=i+1}^{N} S_{i,j}
\end{equation}
where $\mathrm{sgn}(x)=2\theta(x)-1$, and $S_{i,j}\equiv
(1/2)\,\mathrm{sgn}(U_i-U_j)$. 
Obviously $\langle M\rangle=0$ and we
have already calculated the second moment in~\ref{varianceap}, which is
\begin{equation}
\langle M^2 \rangle \equiv\sigma^2= \frac{N(N-1)(2N+5)}{72}
\approx\frac{N^3}{36} \quad\mbox{for large}~N.
\end{equation}

Assuming the pdf of the velocities to be symmetric,
i.e. $\phi(-U)=\phi(U)$, and noting that
$\mathrm{sgn}(-x)=-\mathrm{sgn}(x)$, it can be easily shown that the
probability distribution of $M$ is symmetric about zero. On the other hand,
we have shown in \sref{collision}, that the probability distribution of
$N_c$, and hence that of $M$, is independent of the velocity
distribution. Therefore, the distribution of $M$ must be symmetric for {\em
  any velocity distribution}. This implies the vanishing of all the odd
moments, i.e.,
\begin{equation}
\langle M^{2n+1} \rangle=0, \quad\mbox{for}~ n=0,1,2,\ldots.
\end{equation}
By the same argument, one can also show that average over any product of
odd number of $S_{\{.,.\}}$'s is zero.  $\langle M^{2n+1} \rangle$ is just a sum
of such terms.

The even moments are given by
\begin{equation}
\langle M^{2n} \rangle = 
\sum_{i_1=1}^{N-1}\sum_{j_1=i_1+1}^{N} \cdots
\sum_{i_{2n}=1}^{N-1}\sum_{j_{2n}=i_{2n}+1}^{N} \left\langle 
\prod_{\nu=1}^{2n}S_{i_\nu,j_\nu}
\right\rangle,
\end{equation}
for $n=1,2,3,\ldots$.  We first break the average as product of averages
over $n$ different factors of pair of $S_{\{.,.\}}$'s, and then sum over
the indices independently under each average of factors, each of which
gives the second moment $\langle M^{2} \rangle\equiv \sigma^2$ making it
$\sim \sigma^{2n}$.  This is $\Or(N^{3n})$.  The corrections to this come
from the terms where the sets of indices between different factors are not
independent. If there is one common index between two different factors, we
will have to take the average of them together. It will reduces one
summation and therefore is $\Or(N^{3n-1})$.  The number of ways of grouping
$2n$ objects into $n$ pairs is clearly $(2n)!/[n!\,2^n]$. Therefore,
\begin{equation}
\langle M^{2n} \rangle = \frac{(2n)!}{n!\, 2^n}\;\sigma^{2n}  +
\Or\left(N^{3n-1}\right), \quad\mbox{for}~ n=2,3,\ldots.
\end{equation}

Let us now consider the scaled variable 
\begin{equation}
z=\lim_{N\rightarrow\infty}\; \frac{M}{\sigma}.
\end{equation}
Since $\sigma^{2n}$ is $\Or(N^{3n})$, we have
\begin{equation}
\left\langle z^{2n}\right\rangle = \frac{(2n)!}{n!\, 2^n}\;
~\mbox{and}~
\left\langle z^{2n+1}\right\rangle=0
\,, \quad\mbox{for}~
n=0,1,2,3,\ldots.
\end{equation}
The characteristic function of the probability density of $z$ is then
\begin{eqnarray}
\left\langle\rme^{i \lambda z} \right\rangle &=\sum_{n=0}^\infty 
\frac{(-1)^n}{(2n)!}\; \lambda^{2n}\,\left\langle z^{2n}\right\rangle\nonumber\\
&=\sum_{n=0}^\infty \frac{(-1)^n}{n!}\; \left(\frac{\lambda^2}{2}\right)^n
=\exp\left(-\frac{\lambda^2}{2}\right).
\end{eqnarray}
The inversion of this Fourier transform gives the pdf of $z$ as
\begin{equation}
p(z)=\frac{1}{\sqrt{2\pi}}\; \exp\left(-\frac{z^2}{2}\right),
\end{equation}
which says that the scaled variable $[N_c-\langle N_c\rangle]/\sigma$ has a
Gaussian distribution.

\section{Evaluation of the terms in the series expansion of $F_N(T)$}
\label{terms of F_N}

\subsection*{First term:}

Let us compute the average $\left\langle \theta(x_i)\, \theta(x_i/T-v_i)
\right\rangle$, in the first term (counting from the ``zero''-th term,
which equals unity) in \eref{expansion of F_N}.  The pdf of $v_i$ is given
by
\begin{eqnarray}
\rho_i(v)= &\theta(v)\;\frac{N!}{(N-i-1)!\,(i-1)!}\nonumber\\ 
&\times\int_{-\infty}^\infty\rmd V\,
\phi(V)\,
\Bigl[1-\Phi(V)\Bigr]^{N-i-1} \phi(V+v)\,\Bigl[\Phi(V+v)\Bigr]^{i-1},
\label{rho-v}
\end{eqnarray}
in which
\begin{equation}
\Phi(w)=\int_w^\infty \phi(V)\, \rmd V,\qquad \mbox{i.e.,}\quad \rmd
\Phi=-\phi(V)\,\rmd V.
\label{Phi}
\end{equation}
The integrand in \eref{rho-v} merely specifies that $(N-i-1)$ velocities
are smaller than $V$, $(i-1)$ velocities are larger than $(V+v)$, two
velocities are $V$ and $(V+v)$ respectively, such that their difference is
$v$. The combinatorial prefactor gives the number of such arrangements, and
we finally integrate the integrand over all possible values of $V$.  It is
checked that this pdf is normalized, $\displaystyle\int_0^\infty
\rho_i(v)\,\rmd v=1$.  Using \eref{rho-v} we first take the average over
$v_i$, which for all $i=1,2,\dots, N-1$ gives
\begin{eqnarray}\fl
\theta(x_i)\,\left\langle \theta\Bigl(\frac{x_i}{T}-v_i\Bigr)
\right\rangle_{v_i} =\frac{\theta(x_i)\;N!}{(N-i-1)!\,(i-1)!}  
&\int_{-\infty}^\infty  \rmd V\,
\phi(V)\, \Bigl[1-\Phi(V)\Bigr]^{N-i-1} \nonumber\\
&\times\frac{1}{i}\left(\Bigl[\Phi(V)\Bigr]^i
-\Bigl[\Phi(V+x_i/T)\Bigr]^i\right).
\label{average-v}
\end{eqnarray}
When $T \gg N$,
\begin{equation}
\frac{1}{i}\left(\Bigl[\Phi(V)\Bigr]^i
-\Bigl[\Phi(V+x_i/T)\Bigr]^i\right)=\frac{x_i}{T} \Bigl[\Phi(V)
  \Bigr]^{i-1} \phi(V) +\Or \left(\frac{i x_i^2}{T^2}\right).
\label{taylor1}
\end{equation}
Note that the pdf of $x_i$ as given by \eref{expressiong} is independent of
the index $i$. Substituting \eref{taylor1} in \eref{average-v}, then taking
average over $x_i$, and finally summing over $i$ yields
\begin{equation}
\sum_{i=1}^{N-1}\left\langle \theta(x_i)\, 
\theta\Bigl(\frac{x_i}{T}-v_i\Bigr) \right\rangle
=b\, \frac{N (N-1)}{T} +\Or\left(\frac{N^3}{T^2}\right),
\label{term1}
\end{equation} 
with $b$ given by the \eref{expression of b}. 

\Eref{term1} suggests that $T$ scales as $N^2$, which indicates that the
natural scaling limit corresponds to $T\rightarrow\infty$ and
$N\rightarrow\infty$, while keeping $T/N^2$ fixed. In this limit the
term~\eref{term1} becomes
\begin{equation}
\lim_{N\rightarrow\infty}\sum_{i=1}^{N-1}\left\langle \theta(x_i)\, 
\theta\Bigl(\frac{x_i}{N^2\tau'}-v_i\Bigr) \right\rangle
=\frac{b}{\tau'}.
\label{term1.1}
\end{equation}

\subsection*{Second term:}
Let us now consider the second term in \eref{expansion of F_N} involving
summation over $i$ and $j$. We substitute $T=N^2\tau'$, and rewrite it as
\begin{eqnarray}
\label{1st}
&\sum_{i=1}^{N-3}\sum_{j=i+2}^{N-1}\left\langle \theta(x_i)\theta(x_j)\,  
\theta\Bigl(\frac{x_i}{N^2\tau'}-v_i\Bigr) 
\theta\Bigl(\frac{x_j}{N^2\tau'}-v_j\Bigr) \right\rangle 
\\
\label{2nd}
&+
\sum_{i=1}^{N-2}\left\langle \theta(x_i)\theta(x_{i+1})\,  
\theta\Bigl(\frac{x_i}{N^2\tau'}-v_i\Bigr) 
\theta\Bigl(\frac{x_{i+1}}{N^2\tau'}-v_{i+1}\Bigr) \right\rangle.
\end{eqnarray}
Now in the first term \eref{1st}, the variables $x_i$ and $x_j$ are
uncorrelated, and each of them has the same pdf~\eref{expressiong}.
Following the similar reasoning that we used to write down the pdf of the
single variable $V_i$ in \eref{rho-v}, we can also write down the joint pdf
of $v_i$ and $v_j$ for $j\geqslant i+2$ as
\begin{eqnarray}\fl\qquad
\rho_{i,j} (v_i,v_j)&=\theta(v_i)\,\theta(v_j)\; \frac{N!}{(N-j-1)!\,
  (j-i-2)!\,(i-1)!} \nonumber\\
&\times\int_{-\infty}^\infty\rmd V\,\phi(V) \int_{-\infty}^\infty\rmd V'
\,\phi(V')\,
 \;\theta(V-V')\;\Bigl[1-\Phi(V')\Bigr]^{N-j-1}\nonumber\\
&\rule{3.8cm}{0pt}
\cdot
\theta(V-V'-v_j)\;\phi(V+v_i)\,\phi(V'+v_j)\nonumber\\
&\rule{3.8cm}{0pt}
\cdot
\Bigl[\Phi(V'+v_j)-\Phi(V)\Bigr]^{j-i-2}
\Bigl[\Phi(V+v_i)\Bigr]^{i-1},
\label{rho-vv}
\end{eqnarray}
where $\Phi(w)$ is given by \eref{Phi}. It is checked that the joint pdf is
normalized to unity, i.e., $\displaystyle\int_0^\infty\rmd
v_i\int_0^\infty\rmd v_j\, \rho_{i,j} (v_i,v_j) =1$. Now using
\eref{rho-vv} we first compute the average over $v_i$ and $v_j$ in
\eref{1st}, then we expand the result in Taylor series (assuming large
$N$), and then average over $x_i$ and $x_j$. Finally, we sum over $i$ and
$j$, and take the limit $N\rightarrow \infty$. Noting that
\begin{equation}
\int_{-\infty}^\infty\rmd V\,\phi^2(V) \int_{-\infty}^\infty\rmd V'
\,\phi^2(V')\,
\;\theta(V-V')=\frac{1}{2!}\left[\int_{-\infty}^\infty\phi^2(V)\, \rmd
  V\right]^2 ,
\end{equation}
we find
\begin{equation}\fl\qquad
\lim_{N\rightarrow\infty}
\sum_{i=1}^{N-3}\sum_{j=i+2}^{N-1}\left\langle \theta(x_i)\theta(x_j)\,  
\theta\Bigl(\frac{x_i}{N^2\tau'}-v_i\Bigr) 
\theta\Bigl(\frac{x_j}{N^2\tau'}-v_j\Bigr) \right\rangle =\frac{1}{2!}\; 
\left(\frac{b}{\tau'}\right)^2\,,
\end{equation}
where $b$ is given by \eref{expression of b}.  The term \eref{2nd} goes to
zero in the limit $N\rightarrow\infty$.

\subsection*{The $n$-th term:}
Following the exactly same steps, we can evaluate the $n$-th term of the
expansion of $F_N(T)$ in the expression~\eref{expansion of F_N}.  We first
assume that $n\ll N$, and later take the limit $N\rightarrow\infty$.  We
again write the sum as
\begin{eqnarray}
\label{1st-n}
\sum_{i_1=1}^{N-2n+1}\;
\sum_{i_2=i_1+2}^{N-2n+3} 
\cdots\sum_{i_n=i_{n-1}+2}^{N-1}
\left\langle
\prod_{\nu=1}^n\Biggl[ 
\theta(x_{i_\nu})\, 
\theta\Bigl(\frac{x_{i_\nu}}{N^2\,\tau'}-v_{i_\nu}\Bigr)\Biggr]
 \right\rangle\\
\label{2nd-n}
+\Bigl[\mbox{remaining terms}\Bigr],
\end{eqnarray}
such that \eref{1st-n} contains only the terms with $i_k\geqslant i_{k-1}+2$ for
all $k=2,\ldots,n$. As before, the variables $\{x_{i_\nu},
\nu=1,2,\ldots,n\}$ in \eref{1st-n} are uncorrelated, and the joint pdf of
$\{v_{i_\nu}\}$ can be written as
\begin{eqnarray}\fl
\rho_{i_1,i_2,\ldots,i_n}(v_{i_1},v_{i_2},\ldots,v_{i_n})&=
\left[\prod_{\nu=1}^n \theta(v_{i_\nu})\right]
C_N(i_1,i_2,\ldots,i_n)\nonumber\\
&\times\int_{-\infty}^\infty\rmd V_{(1)}\, \phi\left(V_{(1)}\right)
\int_{-\infty}^\infty\rmd V_{(2)}\, \phi\left(V_{(2)}\right)
\cdots
\int_{-\infty}^\infty\rmd V_{(n)}\, \phi\left(V_{(n)}\right)\nonumber\\
&\rule{1cm}{0pt}\cdot
\left[\prod_{k=1}^{n-1}\theta \left(V_{(k)}-V_{(k+1)}\right)\right]
\cdot
\Bigl[1-\Phi\left(V_{(n)}\right)\Bigr]^{N-i_n-1}
\nonumber\\
&\rule{1cm}{0pt}\cdot
\left[\prod_{k=2}^{n}\theta \left(V_{(k-1)}-V_{(k)}-v_{i_k}\right)\right]
\cdot
\left[\prod_{k=1}^{n} \phi\left(V_{(k)}+v_{i_k}\right)\right]\nonumber\\
&\rule{1cm}{0pt}\cdot
\left[\prod_{k=2}^n\Bigl[\Phi\left(V_{(k)}+v_{i_k}\right)-
\Phi\left(V_{(k-1)}\right) \Bigr]^{i_k-i_{k-1}-2} \right]\nonumber\\
&\rule{1cm}{0pt}\cdot
\Bigl[\Phi\left(V_{(1)}+v_{i_1}\right) \Bigr]^{i_{1}-1},
\label{rho-vv...v}
\end{eqnarray}
where $\Phi(w)$ is given by \eref{Phi} and where the combinatorial
prefactor $C_N$ gives the number of ways of arranging $N$ velocities as in
the above integrand,
\begin{equation}\fl\qquad\quad
C_N(i_1,i_2,\ldots,i_n)=N!\,
\left\{(i_1-1)!\,
\prod_{k=2}^n\Bigl[(i_k-i_{k-1}-2)!\Bigr]\,
(N-i_n-1)!\right\}^{-1}.
\end{equation}
It is again checked that the joint pdf \eref{rho-vv...v} is normalized,
i.e.,
\begin{equation}
\int_0^\infty\cdots\int_0^\infty
\rho_{i_1,i_2,\ldots,i_n}(v_{i_1},v_{i_2},\ldots,v_{i_n})\,
\prod_{\nu=1}^n\rmd v_{i_\nu}=1.
\end{equation}

Using \eref{rho-vv...v} we first take the average over
$\{v_{i_\nu}\}$ in \eref{1st-n}, then we expand the result in Taylor series
(assuming large $N$), and then average over $\{x_{i_\nu}\}$. Finally,
summing over all the indices $\{i_{\nu}\}$, and taking the limit
$N\rightarrow \infty$, \eref{1st-n} yields
\begin{eqnarray}
\label{pre-Int}
\frac{1}{(\tau')^n}\;&\times\left[\int_0^\infty x\, g(x)\,\rmd x \right]^n\\
\label{MInt}
&\times
\int_{-\infty}^\infty\cdots\int_{-\infty}^\infty
\left[\prod_{k=1}^n \rmd V_{(k)}\, \phi^2\left(V_{(k)}\right)\right]\cdot
\left[\prod_{k=1}^{n-1}\theta \left(V_{(k)}-V_{(k+1)}\right)\right].
\end{eqnarray}
Now, the term inside the first square brackets of the integrand in
\eref{MInt} remains unchanged under permutations of the variables
$\{V_{(k)}\}$.  While the term inside the second square brackets of the
integrand changes under permutations, summing over all possible
permutations yields unity.  However, note that $\{V_{(k)}\}$'s are just
dummy variables in the multiple integral \eref{MInt}, so it must remain
unchanged under any of the $n!$ permutations of these variables.
Therefore, the integral \eref{MInt} equals
\begin{equation}
\frac{1}{n!}\int_{-\infty}^\infty\cdots\int_{-\infty}^\infty
\left[\prod_{k=1}^n \rmd V_{(k)}\, \phi^2\left(V_{(k)}\right)\right]
=\frac{1}{n!}
\left[\int_{-\infty}^\infty \phi^2(V)\,\rmd V \right]^n.
\end{equation}
Therefore, in the limit $N\rightarrow\infty$ the expression \eref{1st-n}
becomes [see \eref{pre-Int}, \eref{MInt}],
\begin{equation}\fl\quad
\lim_{N\rightarrow\infty} \sum_{i_1=1}^{N-2n+1}\; \sum_{i_2=i_1+2}^{N-2n+3}
\cdots\sum_{i_n=i_{n-1}+2}^{N-1} \left\langle \prod_{\nu=1}^n\Biggl[
  \theta(x_{i_\nu})\,
  \theta\Bigl(\frac{x_{i_\nu}}{N^2\tau'}-v_{i_\nu}\Bigr)\Biggr]
\right\rangle =\frac{1}{n!}\; \left(\frac{b}{\tau'}\right)^n\,,
\end{equation}
where $b$ is given by \eref{expression of b}.  The remaining term
\eref{2nd-n} goes to zero in the limit $N\rightarrow\infty$.

\addcontentsline{toc}{section}{\protect\numberline{}\refname}


\section*{References}

\begin{thebibliography}{10}

\bibitem{frisch} Frisch H L 1956 Poincar\'e Recurrences {\em Phys. Rev.}
  {\bf 104} 1

\bibitem{teramoto} Teramoto E and Suzuki C 1955 The Statistical Mechanical
  Aspect of H-Theorem {\em Prog. Theor. Phys.} {\bf 14} 411

\bibitem{jepsen} Jepsen D W 1965 Dynamics of a Simple Many-Body System of
  Hard Rods {\em J. Math. Phys.} {\bf 6} 405

\bibitem{lebowitz1} Lebowitz J L and Percus J K 1967 Kinetic Equations and
  Density Expansions: Exactly Solvable One-Dimensional System {\em
    Phys. Rev.} {\bf 155} 122

\bibitem{lebowitz2} Lebowitz J L, Percus J K and Sykes J 1968 Time
  Evolution of the Total Distribution Function of a One-Dimensional System
  of Hard Rods {\em Phys. Rev. E} {\bf 171} 224

\bibitem{lebowitz3} Aizenman M, Lebowitz J L and Marro J 1978
  Time-displaced correlation functions in an infinite one-dimensional
  mixture of hard rods with different diameters {\em J. Stat. Phys.} {\bf
    18} 179

\bibitem{mckean} McKean H P 1967 Chapman-Enskog-Hilbert expansion for a
  class of solutions of the telegraph equation {\em J. Math. Phys.} {\bf 8}
  547

\bibitem{keyes} Protopopescu V and Keyes T 1985 The Goldstein-McKean model
  revisited {\em Physica} A {\bf 132} 421

\bibitem{lieb} Lieb E 1999 Some problems in statistical mechanics that I
  would like to see solved {\em Physica} A {\bf 263} 491

\bibitem{gruber} For a critical discussion of the ``adiabatic piston
  problem" see Gruber Ch 1999 Thermodynamics of systems with internal
  adiabatic constraints: Time evolution of the adiabatic piston {\em
    Eur. J. Phys.} {\bf 20} 259

\bibitem{jarek1} Piasecki J and Gruber Ch 1999 From the adiabatic piston to
  macroscopic motion induced by fluctuations {\em Physica} A {\bf 265} 463

\bibitem{jarek2} Piasecki J and Sinai Ya G 2000 A model of non-equilibrium
  statistical mechanics in {\em Dynamics: Models and Kinetic Methods for
    Non-equilibrium Many Body Systems} Karkheck J (ed) 
    (Amsterdam: Kluwer Academic) pp~191-199 

 
\bibitem{jarek3} Piasecki J 2001 Drift Velocity Induced by Collisions {\em
  J. Stat. Phys.} {\bf 104} 1145

\bibitem{bala1} Balakrishnan V, Bena I and Van den Broeck C 2002 Velocity
  correlations, diffusion, and stochasticity in a one-dimensional system
  {\em Phys. Rev.} E {\bf 65} 031102

\bibitem{bala2} Balakrishnan V and Van den Broeck C 2005 Analytic
  calculation of energy transfer and heat flux in a one-dimensional system
  {\em Phys. Rev.} E {\bf 72} 046141

\bibitem{Jarzynski} Jarzynski C 1997 Nonequilibrium Equality for Free
  Energy Differences {\em Phys. Rev. Lett.} {\bf 78} 2690

\bibitem{ioana} Bena I, Van den Broeck C and Kawai R 2005 Jarzynski
  equality for the Jepsen gas {\em Europhys. Lett.}  {\bf 71} 879

\bibitem{sachdev1} Sachdev S and Damle K 1997 Low Temperature Spin
  Diffusion in the One-Dimensional Quantum $\Or(3)$ Nonlinear $\sigma$ Model
  {\em Phys. Rev. Lett.} {\bf 78} 943

\bibitem{sachdev2} Sachdev S and Young A P 1997 Low Temperature
  Relaxational Dynamics of the Ising Chain in a Transverse Field {\em
    Phys. Rev. Lett.} {\bf 78} 2220


\bibitem{eigen} Eigen M 1971 Selforganization of matter and the evolution
  of biological macromolecules {\em Naturwissenschaften} {\bf 58} 465

\bibitem{krug} Krug J and Karl C 2003 Punctuated evolution for the
  quasispecies model {\em Physica} A {\bf 318} 137

\bibitem{jain1} Jain K and Krug J 2005 Evolutionary trajectories in rugged
  fitness landscapes {\em J. Stat. Mech.: Theor. Exp.}  {\bf P04008}


\bibitem{jain2} Jain K 2007 Evolutionary dynamics of the most populated
  genotype on rugged fitness landscapes {\em Phys. Rev.} E {\bf 76} 031922


\bibitem{satya1} Sire C, Majumdar S N and Dean D S 2006 Exact solution of a
  model of time-dependent evolutionary dynamics in a rugged fitness
  landscape {\em J. Stat. Mech.: Theor. Exp.} {\bf L07001}

\bibitem{ioana07} Bena I and Majumdar S N 2007 Universal extremal
  statistics in a freely expanding Jepsen gas {\em Phys. Rev.} E {\bf 75}
  051103

\bibitem{satya06} Majumdar S N and Martin O C 2006 Statistics of the number
  of minima in a random energy landscape {\em Phys. Rev.} E {\bf 74} 061112

\bibitem{krug1} Krug J 2002 Tempo and mode in quasispecies evolution in {\em
  Biological Evolution and Statistical Physics}  L\"assig M and 
  Valleriani A (eds) (Berlin: Springer) pp~205-216 ({\em Preprint}
  arXiv:cond-mat/0103443)


\bibitem{visco} Visco P, van Wijland F and Trizac E 2008 Collisional
  statistics of the hard-sphere gas {\em Preprint} arXiv:0803.1291

\end{thebibliography}
\end{document}